\documentclass[twocolumn]{revtex4-1}
\usepackage{amsmath}    
\usepackage{graphicx}   
\usepackage{verbatim}   
\usepackage{color}      
\usepackage{subfigure}  
\usepackage{hyperref}   
\usepackage{bm} 
\usepackage{multirow}
\usepackage{csquotes}
\newcommand{\classoption}[1]{\texttt{#1}}

\newcommand{\sqrtSnn}{\ensuremath{\sqrt{s_{\mathrm{NN}}}\text{ = 5.44 TeV}}}
\newcommand{\sqrtE}{\ensuremath{\sqrt{s_{\mathrm{NN}}}}}
\newcommand{\pt}{\ensuremath{p_{\mathrm{T}}}}
\newcommand{\fq}{\ensuremath{F_{\mathrm{q}}}}

\newcommand{\fqtwo}{\ensuremath{F_{2}}}

\expandafter\ifx\csname package@font\endcsname\relax\else
\expandafter\expandafter
\expandafter\usepackage
\expandafter\expandafter
\expandafter{\csname package@font\endcsname}%
\fi
\DeclareRobustCommand\substyle{\name@idx{document substyle}}%
\DeclareRobustCommand\classoption{\name@idx{document class option}}%
\DeclareRobustCommand\classname{\name@idx{document class}}%
\def\name@idx#1#2{
	{\ttfamily#2}%
	\index{#2\space#1=\string\ttt{#2}\space#1}\index{#1>#2=\string\ttt{#2}}%
}
\begin{document}
	\title{ Scaling behaviour of charged particles generated in Xe$-$Xe~collisions at $\sqrt{s_{\rm{NN}}}$~=~5.44~TeV using the AMPT model}
\author{Zarina Banoo}
\author{Ramni Gupta}
\email{ramni.gupta@cern.ch}
\author{Salman Khurshid Malik}
\author{Sheetal Sharma}
\author{Fakhar Ul Haider}
\author{Balwan Singh}
\affiliation{Department of Physics, University of Jammu, Jammu, J\&K, India} 
\date{\today}
 \begin{abstract}
 The spatial configurations of particles produced in the kinematic phase space during a heavy-ion collision reflect the characteristics of the system created in the collision. The scaling behaviour of the multiplicity fluctuations is studied for the charged particles generated in Xe--Xe collisions at $\sqrt{s_{\rm{NN}}}$~=~5.44~TeV using the String Melting (SM) mode of the AMPT (A Multi-Phase Transport) model. The scaling behaviour of the normalized factorial moments ($F_\text{q}$) gives significant information about the dynamics of the system under study. A linear power-law growth of the $F_\text{q}$ with the increasing phase space resolution, termed as intermittency, is investigated. The anomalous fractal dimension $D_\text{q}$ is determined, which is linked to the self-similarity and fractal nature of the particle emission spectra, whose  dependence on the order of the moment ($q$) is characterised by the intermittency index ($\varphi_{\text{q}}$). Relating $q^{\rm{th}}$ order Normalised Factorial Moment (NFM) with $F_{2}$, the scaling exponent ($\nu$) is determined that quantifies the dynamics of the system created by these collisions and is analyzed for its dependence on the transverse momentum bin width ($\Delta p_\text{T}$). Results presented may be interpreted as model predictions and baseline expectations.
 \end{abstract} 
\maketitle
\large
\linespread{1.5}
\label{key}	\section{Introduction}	
Identifying the nature of the phase transition and understanding the mechanism leading to multiparticle production in the heavy-ion collisions are among the main objectives of present-day high energy physics experiments. Two distinct types of phase transitions occur when two nuclei collide. The first one is due to the compression of the nuclei under high energy and density, which results in the deconfinement of the quarks and is known as the hadron-quark phase transition and the second is the quark-hadron phase transition, which occurs when the Quark-Gluon Plasma (QGP) expands sufficiently and the confinement forces set in at the end of the evolution to form hadrons~\cite{Hwa:2014xxa}. According to the Quantum Chromodynamics (QCD) theory of strongly interacting matter, the phase transition across the critical line may occur for large values of baryon chemical potential ($\mu_{\text{B}}$) with a Critical End Point (CEP) that is expected to exist at the termination of the first-order phase transition line. This endpoint is characterized by a second-order phase transition based on generic principles  whereas when $\mu_{\text{B}}= 0$, the transition transforms into a smooth crossover~\cite{Stefanek:2014vva, Antoniou:2009abc}.   
\par
During the process of phase transition, conventional Quantum Electrodynamic (QED) matter exhibits large-scale fluctuations as the system approaches the critical point, characterized by a divergence in correlation length. This phenomenon manifests as critical opalescence -- a classical signature of criticality in thermodynamic systems~\cite{Luo:2017zza}. Analogously, in QCD matter created in high-energy heavy-ion collisions, a similar rise in correlation length is expected near the CEP, leading to the development of significant number density fluctuations during the phase transition. These fluctuations span multiple length scales, resulting in spatial patterns that exhibit features of self-similarity or critical clustering. As such, observables related to multiplicity fluctuations can serve as effective signatures of critical phenomena, capable of revealing the presence of patches of all sizes in the produced particle distributions~\cite{Luo:2017zza, Yang:2017aaz}. The investigation of multiplicity fluctuations is therefore a powerful tool for detecting the formation of QGP~\cite{Mukherjee:2016yya}, gaining insight into the mechanisms of particle production, determining the location of CEP, and probing the order of the QCD phase diagram.
\par
Intermittency studies analyze fluctuations in the number density of particles in a phase space. Experimental investigations have observed intermittent patterns, e.g., in hadron-nucleon~\cite{Holynski:1989,Holynski:1989aa}, muon-hadron~\cite{Derado:1990,Zhong:1994} and leptonic interactions~\cite{Parashar:1996xmv, Feindt:1990zf} but no conclusive evidence about the mechanism responsible for the multiparticle production and about critical fluctuations is obtained. Recently, researchers have proposed to study this data-hungry methodology for high multiplicity events from the LHC~\cite{Hwa:2014xxa, Yang:2017aaz}. An analysis on similar lines using Toy MC is reported in~\cite{Sharma:2023oxo}. The AMPT (A Multi Phase Transport) model is capable of providing kinetic description of the main stages of the high-energy collisions and recent studies have highlighted its robust capability to elucidate particle production mechanism in hadron-nucleus and nucleus-nucleus collisions with energy of collisions ranging from 5 to 5500~GeV~\cite{Lin:2005vza}. Scaling behaviours of the fluctuations in the multiplicity distributions obtained from the collisions between xenon (Xe) nuclei using the string melting mode of the AMPT model are investigated.
\par
The article is divided into five sections. In the next section, a brief overview of the AMPT model is given, followed by the analysis methodology in section 3. Observations and results from the analysis of the AMPT events for the study of scaling behaviour of the charged particle generation in Xe--Xe system are discussed in  section 4 followed by a summary in section 5.
\section{The AMPT Model}
A hybrid transport Monte Carlo model, the AMPT model simulates the heavy-ion collisions at ultra-relativistic energies~\cite{Lin:2005vza,Lin:2011aca}. It consists of four subprocesses: the initialization of the collision, partonic transport, hadronization (i.e., conversion from the partonic to the hadronic phase), and finally, hadronic transport~\cite{Zhang:1999vza}. As in all models the outcome particle distributions are dependent on the input parameters which include initial partonic phase and final hadronic phase. The two modes of the AMPT model, the Default (DF) mode and the String Melting (SM) mode, differ in the hadronization process. The initial conditions are taken from the HIJING (Heavy Ion Jet Interaction Generator)~\cite{Wang:1991abc} event generator that gives the spatial and the momentum distributions of minijet partons accompanied by soft string excitations. Radiations originating from various states of HIJING are kept off. The partonic interactions are simulated using Zhang's Parton Cascade (ZPC) model~\cite{Lin:2005vza}, which emphasizes two-body interactions for which the cross sections were obtained based on perturbative Quantum Chromodynamics (pQCD).
\par
After the partonic phase ends, these partons recombine with their parent strings,  which then fragment into hadrons using the Lund string fragmentation model \cite{Andersson1983}. In contrast, in the SM mode, all excited strings from HIJING first fragment into hadrons using the LUND fragmentation and then convert to partons according to their flavour and spin~\cite{Lin2002} before the parton cascade begins. These partons, along with the minijet partons, undergo scatterings via ZPC. After the partonic interactions cease, hadronization is carried out using a quark coalescence model. Thus the SM mode offers an improved description of collective behaviour and multi-particle correlations compared to the default mode, where particle production is largely governed by string fragmentation. In both modes, the subsequent hadronic interactions are modeled using the A Relativistic Transport (ART) model~\cite{Lin:2005vza}, which has been extended to include additional reaction channels relevant at higher energies. The hadronic phase continues until kinetic freeze-out, beyond which further interactions are assumed not to affect the final observables ~\cite{Lin:2011aca}.
 \begin{figure}[t!]
	\centering
	{
    \includegraphics[width=9.0cm, height=7.5cm]{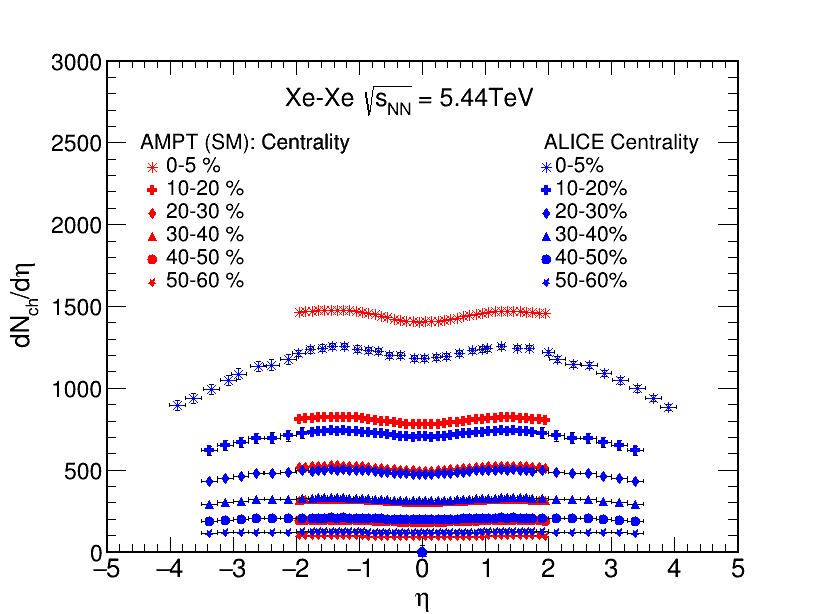}}
	\caption{A comparison of SM AMPT model and the ALICE data~\cite{ALICE:2018collab} for their pseudorapidity density distributions of the charged particles generated in the different centrality Xe--Xe collisions at $\sqrt{s_{\rm{NN}}}$~=~5.44 TeV.}
	\label{fig:figure1}
\end{figure}
\begin{figure}[t!]
	\centering
	{
	\includegraphics[width=9.2cm, height=6.7cm]{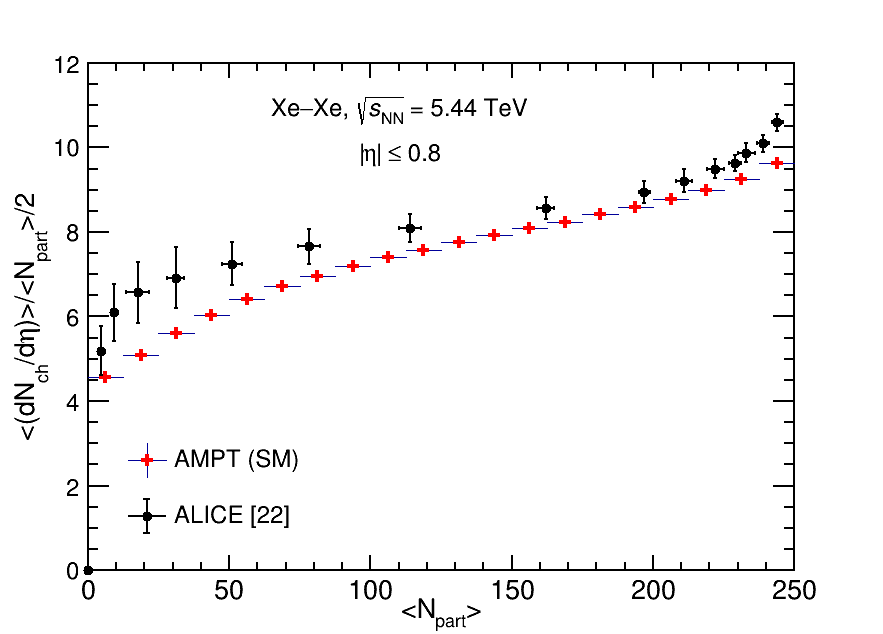}}
	\caption{
	    Normalised charged particle density distributions as a function of  the number of participants from the AMPT (SM) model and the ALICE data~\cite{ALICE:2018collab}, for Xe--Xe~collisions at $\sqrt{s_{\rm{NN}}}$~=~5.44 TeV.
    }
	\label{fig:figure2} 	
\end{figure}
\par
  For the analysis presented here, the event samples are generated using the SM mode of the AMPT (\textit{Ampt-v1.26t9b-v2.26t9b}) model. The Lund string fragmentation parameters are set to $a$ = 0.55 and $b$ = 0.15 GeV$^{-2}\,$, following the standard tuning used in the AMPT model to reproduce charged-particle multiplicities and bulk particle production in relativistic heavy-ion collisions \cite{{Lin2005}}. These parameters primarily control the longitudinal momentum distribution and transverse mass suppression in the Lund fragmentation function and are therefore constrained by global observables such as multiplicity and transverse momentum ($p_{\rm{T}}$) spectra. Furthermore, the parton cross-section is adjusted by setting the QCD coupling constant $\alpha_s$ = 0.33 and the screening mass $\mu$ = 2.256 fm for gluons in the QGP. For these parameters, events for different centrality classes are generated for Xe--Xe collisions at $\sqrt{s_{\rm{NN}}}$~=~5.44~TeV where the centrality percentile values of the impact parameter are taken from Ref.~\cite{CERN:Enterria2018}. The density distributions of the charged particles in the pseudorapidity interval of $|\eta|\le 2.0$, for different collision centrality ranges, are compared with those from the ALICE data points~\cite{ALICE:2018collab} for the same centrality percentiles as are shown in Fig.~\ref{fig:figure1}. Also the normalized charged particle density distributions as a function of number of participants are shown in Fig.~\ref{fig:figure2}. These figures show that the event sample generated with the AMPT model has qualitative agreement with the experimental data from the ALICE~\cite{ALICE:2018collab} experiment at LHC and is a suitable candidate for further analysis to investigate the scaling behaviour of the charged particle generation so as to get an insight into the baseline particle production mechanism and the order of the quark-hadron phase transition.           
\section{Analysis Methodology}
\label{methodolgy}
In ultra-relativistic heavy-ion collisions with high particle number density per bin in the phase space, the event-by-event study of particle multiplicity distributions, to extract the signal of interest, is favoured~\cite{Jeon:2003nza}. The pseudorapidity ($\eta$) and azimuth angle ($\phi$) define the angular distributions of particles produced in these collisions~\cite{PDG2022}. The spatial configurations of the particles produced depend on the dynamics of the system created in the collisions. With different sets of collision conditions these configurations may vary from event-to-event. The scaling behaviour of the moments of these particle number density distributions (the spatial configurations) is one of the important characteristics to understand the system's dynamics.
\par 
A two-dimensional intermittency analysis technique as proposed in~\cite{Hwa:2011bu} is used to calculate normalized factorial moments (NFM) that gauge bin-to-bin fluctuations in particles generated in ($ \eta,\phi$) phase space of an event. The phase space is partitioned into a lattice of cells with $M$  number of bins along each dimension so as to have a square lattice with a total of M$_{\eta} \times {M_{\phi}}$ bins. M$_{\eta}$ and M$_{\phi}$ being the number of bins along $\eta$ and $\phi$ dimension respectively. The particles generated in an event are mapped onto this lattice that may result in different number of particles per cell. The distribution of particles in the bins or cells gives spatial configuration of particles in the phase space. Varying the number of partitions (M), the number of particles that go into each cell (known as bin multiplicity) changes. If $n_{\rm{m}}$ defines the bin multiplicity in the $m^{\rm{th}}$ bin of an $e^{\rm{th}}$ event then the event factorial moment of order $q$ with condition that bin multiplicity $n_m \ge \textit{q}$ is defined as  
\begin{equation}
\begin{split}
f_q^{e}(M)\,=\,<n_m(n_m-1)\dots(n_m-q+1)>.
\end{split}
 \label{eqn:01}
\end{equation}  
Here $<\ldots>$ is the averaging over the total number of bins ($M^{\rm{2}}$,  for ${M_{\eta}}$ = $M_{\phi}$ = M (say)) in the two-dimensional phase space. Then for a sample of N events, the NFM $F_{\rm{q}}(M)$ is defined as~\cite{Hwa:2011bu};
\begin{equation}
\begin{split}
F_q(M)=\frac{1}{N }\sum_{e=1}^{N}\frac{<n_m(n_m-1)\dots(n_m-q+1)>_e}{[<n_m>_{e}]^{q}},
\end{split}
 \label{eqn:1}
\end{equation}  
 where $\textit{q}$, the order of the moments, is a positive integer $\ge 2$. These normalized factorial moments have the property of detecting and characterizing the dynamical fluctuations indicated by the dependence of $F_{\rm{q}}(M)$ on the decreasing phase space cell size ($\delta$) or increasing number of bins (as M~$\propto$ 1/$\delta$), such that
\begin{eqnarray}
F_q(\delta) \propto (\delta)^{-\varphi_{\rm{q}}},
\label{eqn:2}
\end{eqnarray}
where $\varphi_{\rm{q}}$ is known as the intermittency index and this power-law scaling behaviour is termed as \textit{intermittency}~\cite{Wolf:1996aan}. The intermittency study of heavy-ion data has been successfully explained to study the high spike event observed in the cosmic ray data in the JACEE experiment~\cite{Burnett:1983abc}. In terms of number of bins in the phase space, Eq.\eqref{eqn:2} can be written as,
\begin{eqnarray}
F_q(M) \,\propto\, (M^{d})^{\varphi_{\rm{q}}}.
\label{eqn:Mscaling}
\end{eqnarray}
In the above equation $d$ defines the number of dimensions of the phase space. Scaling of $F_{\rm{q}}(M)$ with the number of bins as in Eq.\eqref{eqn:Mscaling} is also known as \textit{M-scaling} or resolution scaling. \\
$\varphi_{q}$ is related  to the anomalous fractal dimension $d_{\rm{q}}$ relating it to the order of the moment $\textit{q}$ as~\cite{{Sarkisian:1993pi}}. 
  \begin{equation}
  d_q = \frac{\varphi_q}{(q-1)}.
  \label{eqn:dq}
  \end{equation}
The significance of anomalous fractal dimension is that it reveals the nature of the fractal structure of the system under study. Further $d_{\rm{q}}$ is related to the generalized fractal dimension $(D_{\rm{q}})$ through the relation~\cite{Sarkisian:1993ne,Gelovani:1997ne},
\begin{equation}
   d_{q} = D_{\rm{T}}-D_{\rm{q}},
   \label{eqn:dq1}
\end{equation}
where $D_{T}$ and $D_{q}$ are ordinary topological dimension and generalized (or Renyi) fractal dimension~\cite{Lipa:1989aaz} respectively. A distinct characterization of fractal behaviour is observed based on the dependence of $d_{\rm{q}}$ on the moment order $q$. A system with \textit{multifractal} nature~\cite{Gupta:2019zox} shows dependence of $d_{\rm{q}}$ on \textit{q}, and \textit{monofractal} nature in case there is no such dependence. 
\begin{figure}[tp!]
 	\centering
 	\includegraphics[width=8.8cm, height=6.5cm]{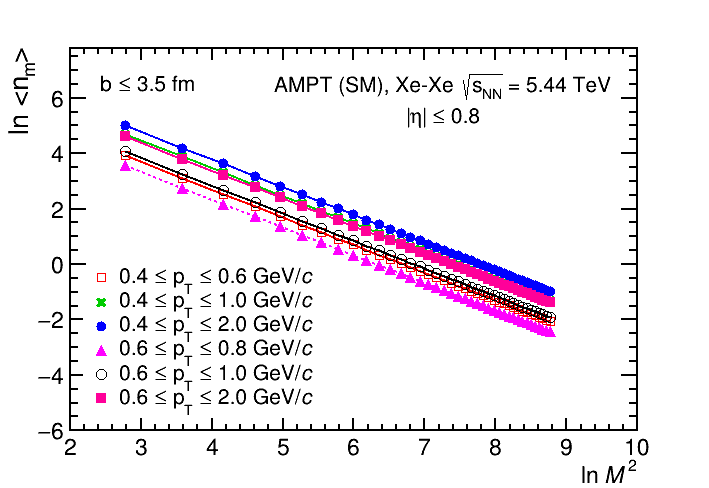}
 	\caption{Average bin multiplicities of the charged particles generated in the mid rapidity region in the various $p_{\rm{T}}$ intervals below 2.0 GeV/\textit{c} as a function of number of bins.}
 	\label{fig:figure1a}
\end{figure}  
  \begin{figure}[t!]
 	\includegraphics[width=8.8cm, height=6.4cm]{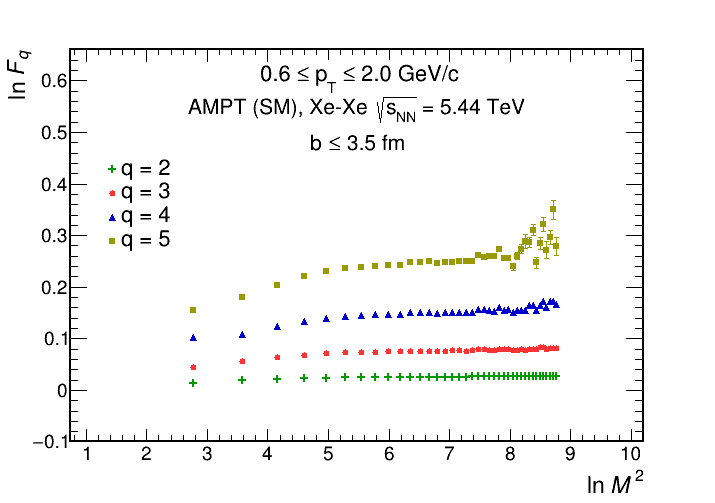}
    \caption{log-log plot of the NFM $F_{\rm{q}}(M)$ for $q$ = 2, 3, 4 and 5  as a function of the number of cells ($M^{2}$) for the $p_{\rm{T}}$ interval $0.6 \leq p_\text{T} \leq 2.0$~GeV/\textit{c}. Similar observations are made for the other transverse momentum bins.}
   	\label{fig:figure51}
  \end{figure} 
\par
Bialas and Zalewski ~\cite{Bialas:1990vh}  suggested  that phase transition, occurring in systems far away from the thermal equilibrium, could be the key to the emergence of intermittent behaviour~\cite{Peschanski:1989yu, Peschanski2:1991}. It is proposed that particle distributions across various spatial and temporal scales can be characterized by a coefficient, $\lambda_{\rm{q}}$ related to the intermittency index $\varphi_{\rm{q}}$ through \textit{q} as
   \begin{equation}
  \lambda_q = \frac{(\varphi_q + 1)}{q}.
  \label{eqn:lambda}
  \end{equation}
For the non-thermal phase transition  $\lambda_{\rm{q}}$ is predicted to go through a minimum at some value of $q=q_{\rm{c}}$~\cite{Ahmad:2010aaz} which separates the region with large number of bins containing particles $\leq q_{\rm{c}}$  and region ($q \geq q_{\rm{c}}$) with a few bins having rare large-multiplicity spikes $\geq q_{\rm{c}}$. Identifying the order parameter with the multiplicity density, to a high degree of accuracy, it was observed~\cite{Hwa:1992uq} that $F_{\rm{q}}$ satisfies the power-law behaviour
\begin{align}
F_q(M) \propto F_{2}(M)^{\rm{\beta_{\rm{q}}}},
\label{eqn:fscaling}
\end{align}
where $\beta_{\rm{q}}$ = $\varphi_{\rm{q}}$/$\varphi_{\rm{2}}$. The scaling of higher order NFM with the second order NFM as in the above equation is termed as \textit{F-scaling} or \textit{order scaling}. $\beta_{\rm{q}}$ is observed to be related to $q$ through a scaling exponent $\nu$ as;
\begin{equation}
\beta_{q} = (q-1)^\nu.
\label{eqn:Mscaling2}
\end{equation}
The scaling exponent $\nu$ is a dimensionless parameter that characterizes the system being studied. It quantifies the fluctuations in the spatial patterns of the particles generated in a collision and provides a quantitative measure of the underlying dynamics. Being independent of units, the characteristics of various systems from experiments~\cite{Yang:2017aaz} and models can be compared using this scaling exponent. From the Ginzburg-Landau theory with formalism for second order phase transition, $\nu$ is found to have an average value of 1.304~\cite{Hwa:1992uq, Ochs:1988ky}. From the two-dimensional Ising model, intermittency analysis gives average value of $\nu$ $\approx 1.3$~\cite{Hwa:1992uq}.  With critical fluctuations introduced in Ising model $\nu$~=~1.04. The scaling exponent value obtained from Pb--Pb collisions at $\sqrtE=$ 2.76 TeV using the SM AMPT model is $\sim 1.7$ \cite{Sharma:2018vtf}. However, the Successive Contraction and Randomization (SCR) model containing fluctuations of critical nature gives $\nu$~=~1.41~\cite{Yang:2017aaz}.
\par
In the following section observations and results from this analysis performed on the Xe--Xe collision events generated using the SM AMPT model are presented and discussed. Similar analyses within the AMPT framework have been reported for Pb--Pb collisions at $\sqrt{\rm{s_{NN}}}$ = 2.76 TeV \cite{Sharma:2018vtf} and Au--Au collisions at RHIC energies~\cite{Xie2013}. However, to the best of our knowledge, a detailed study of intermittency behaviour in Xe--Xe collisions at $\sqrtSnn$ within the SM mode of the AMPT model has not been reported so far. The Xe--Xe system represents an intermediate-sized collision system, providing a unique opportunity to investigate the system-size dependence of multiplicity fluctuations and scaling behaviour within a consistent framework. In particular, the deformed nuclear geometry of Xe nuclei and the corresponding initial-state conditions generated during the collision can significantly influence particle production dynamics and fluctuation patterns. The present work therefore investigates intermittency and scaling behaviour in Xe--Xe collisions within the SM AMPT framework, with the objective of establishing baseline expectations for the observables and examining how nuclear deformation and collision initial conditions may influence the emergence of scaling behaviour through the underlying transport dynamics.
\begin{figure}[h!]
	\centering
	\includegraphics[width=9.3cm, height=6.6cm]{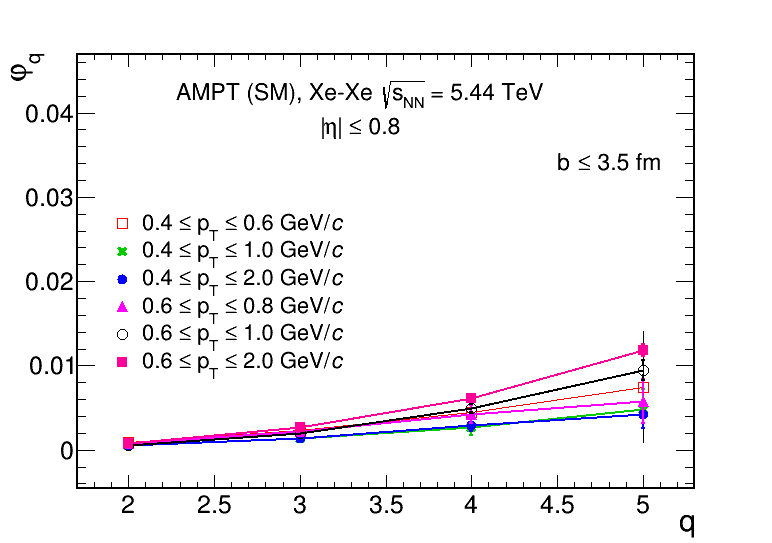}  	
	\caption{Intermittency indices $(\varphi_{\rm{q}})$ as a function of order of the moments \textit{q}~=~2,~3,~4,~5.}
	\label{fig:figure52}
\end{figure}
\section{Observations and Results }
The charged particle multiplicity distributions obtained from Xe--Xe collisions at $\sqrtSnn$ using the SM mode of AMPT model are investigated using the normalized factorial moments ($F_{\rm{q}}$). Around half a million minimum bias events are generated. Of these around  $3\times10^5$ central events having impact parameter $b \leq 3.5$ \textit{fm} are analysed to study scaling behaviour of the NFM as discussed in the above section. The charged particles (protons, pions, and kaons) generated in the two-dimensional angular phase space with $|\eta| \leq 0.8$ and full azimuthal angle in the soft transverse momentum region are studied to investigate their two dimensional intermittency and fractal behaviour. Two sets of $p_{\rm{T}}$ bins: narrow $p_{\rm{T}}$ bins of width $\Delta p_{\rm{T}}$~=~0.2 ($0.4 \leq p_{\rm{T}} \leq 0.6$ GeV/c and $0.6 \leq p_{\rm{T}} \leq 0.8$ GeV/c) and a few wider $p_{\rm{T}}$ bins ($0.4 \leq p_{\rm{T}} \leq 1.0$ GeV/c, $0.6 \leq p_{\rm{T}} \leq 1.0$ GeV/c, $0.4 \leq p_{\rm{T}} \leq 2.0$ GeV/c and $0.6 \leq p_{\rm{T}} \leq 2.0$ GeV/c) are considered. The charged particles within the kinematic acceptance region are mapped onto $M^{\rm{2}}$ cells in the two dimensional phase space that is partitioned with $M$ bins along each dimension. $M$ takes value from a minimum of 4 to a maximum value of 80. The maximum number of bins along each dimension depends on the charged particle density and in case of experimental study it also depends on the detector resolution. The number of particles that fall in a cell that is, bin multiplicity depends on the dynamics of particle generation. For the charged particles in different $p_{\rm{T}}$ intervals the average bin multiplicity over all events, as a function of the number of bins $M^{\rm{2}}$, is shown in Fig.~\ref{fig:figure1a}. The average bin-multiplicity per bin, $\langle n_m\rangle$, is observed to decrease with increasing M, which is a trivial consequence of subdividing the phase space into a larger number of bins. For a given M, the average bin content is less in the small $p_T$ intervals with  $\Delta p_\text{T}$ = 0.2 compared to wider  $p_{\rm{T}}$ bins, reflecting the reduced particle yield in a smaller $p_{\rm{T}}$ range.  Although these features are expected and do not carry independent physical significance, the figure is presented to illustrate the basic behaviour of the data and to serve as a reference for the subsequent analysis.
\par
The NFM, $F_{\rm{q}}(M)$ are calculated for the order of the moments $q$ = 2,~3,~4 and 5. Fig.~\ref{fig:figure51} shows a case of log-log plot of $F_{\rm{q}}(M)$ versus $M^{\rm{2}}$ for the charged particles in transverse momentum interval $0.6 \leq p_{\text{T}} \leq 2.0$ GeV/{\textit{c}}. The error bars on the markers are the statistical uncertainties calculated using the sub-sampling method. With increase in $q$ value the $F_{\rm{q}}(M)$ values also increase that is $F_{\rm{q+1}}(M) \geq F_{\rm{q}}(M)$. Similar trends have been observed in all transverse momentum intervals. $F_{\rm{q}}$ is observed to grow for the small values of $M$, followed by saturation. This behaviour of $F_{\rm{q}}(M)$ with M shows absence of large fluctuations in the spatial distributions of the particles that is, there are no substantial bin-to-bin fluctuations in the particle generation in the events. In other words there is no intermittency and no scale invariance in the multiplicity fluctuations, an observation that is anticipated because the phase transition physics and self-similar particle generation processes not being incorporated in the SM AMPT model. Thus within the SM AMPT framework, the observations from the analysis provide direct access to the fluctuation patterns arising from the model dynamics. It is pertinent to mention that $F_{\rm{q}}(M)$ calculations are made as per the formalism proposed in~\cite{Hwa:2011bu} using the same-event multiplicity distributions. This approach is complementary to experimental methodologies that may employ additional procedures to account for detector and background effects.
\par
\begin{figure}[h]
  	\centering
  		\includegraphics[width=9.5cm, height=6.8cm]{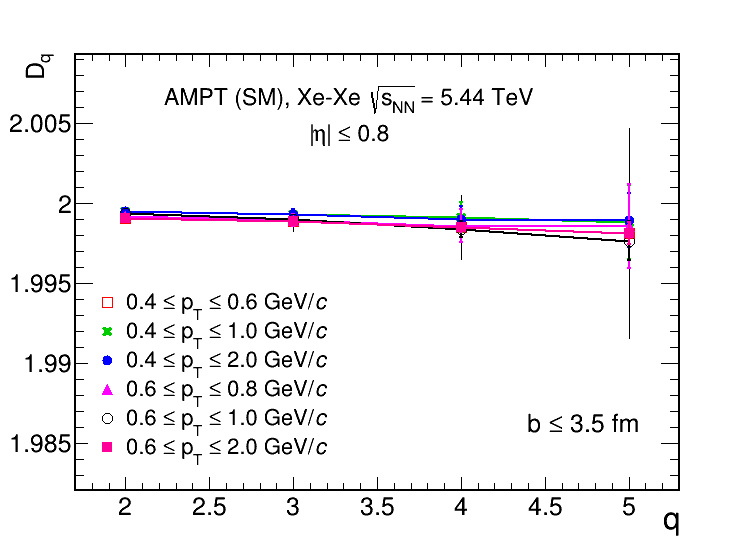}	
  	\caption{The generalized dimension $(D_{\rm{q}})$ as a function order of moments {\textit{q}}  from the intermittency analysis of the charged particles generated in the central SM AMPT events. $D_{\rm{q}}$ is observed to be independent of q. }
  	\label{fig:figure12}
  \end{figure}

The intermittency indices $\varphi_{\rm{q}}$, the slopes from the line fits performed in the high M region of the $\ln F_{\rm{q}}$ vs $\ln M^{\rm{2}}$ plots for $q$~=~2 to 5, are obtained. $\varphi_{\rm{q}}$ as a function of the order of the moments ($q$) in different transverse momentum intervals is shown in Fig.~\ref{fig:figure52}. Error bars on the markers are the fitting errors. It is observed that within errors, $\varphi_{\rm{q}}$ is independent of $q$ further indicating  absence of bin-to-bin fluctuations in the central collision events. As discussed above, the flat behaviour of NFM with $M^{\rm{2}}$, for all $p_{\rm{T}}$ bins is observed (Fig.~\ref{fig:figure51}). Using $\varphi_{\rm{q}}$, fractal dimensions $D_{{\rm{q}}}$ (Eq.(\ref{eqn:dq1})) are  determined for $q$ = 2, 3, 4, 5 and $D_{\rm{T}}=2$. The dependence of $D_{\rm{q}}$ on $q$ for all the $\pt$ bins is shown in Fig.~\ref{fig:figure12}. Particle generation is observed to have \textit{monofractal} nature as $D_{\rm{q}}$ is independent of \textit{q} and $p_{\rm{T}}$ bins as well. Thus no bin-to-bin fluctuations in charged particle generation in the SM AMPT Xe--Xe collision central events. 
\par
\begin{figure}[t]
\centering
   	\includegraphics[width=9.cm,height=6.6cm]{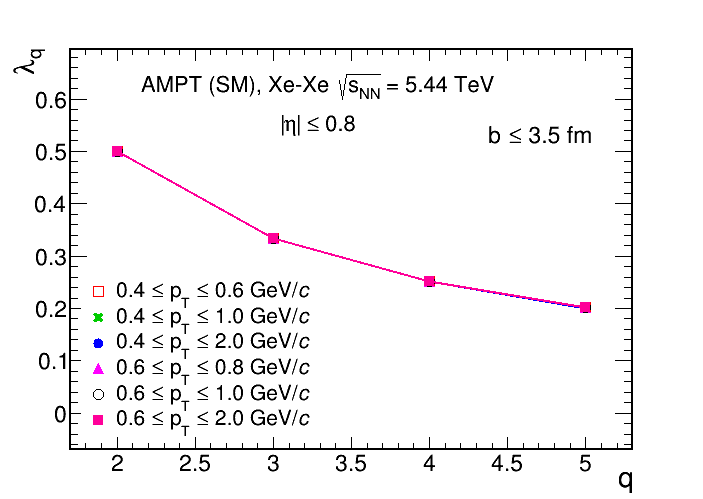}
   	\caption{$\lambda_{\rm{q}}$ as a function of order of moments (q)  for SM AMPT events in the various $p_{\rm{T}}$ bins.}
   	\label{fig:figure13}
\end{figure}

The coefficient $\lambda_{\rm{q}}$ as a function of $q$ (= 2, 3, 4, 5) in the various $p_T$ bins is shown in Fig.~\ref{fig:figure13}. The parameter $\lambda_{\rm{q}}$ exhibits a clear dependence on $q$, unlike $D_{\rm{q}}$, which shows no such variation. Large statistical errors on $F_{\rm{q}}(M)$ for $q \geq$ 5 have limited the study upto $q$~=~5. However, $\lambda_{\rm{q}}$ values are independent of the $p_{\rm{T}}$ bins studied here. This means that the particle generation mechanism for all the $p_{\rm{T}}$ values in the SM AMPT is similar. As of now the existence of two phase system formation has not been observed in any experimental study.\\

    \begin{figure}[t]
   	\includegraphics[width=9.0cm, height=6.2cm]{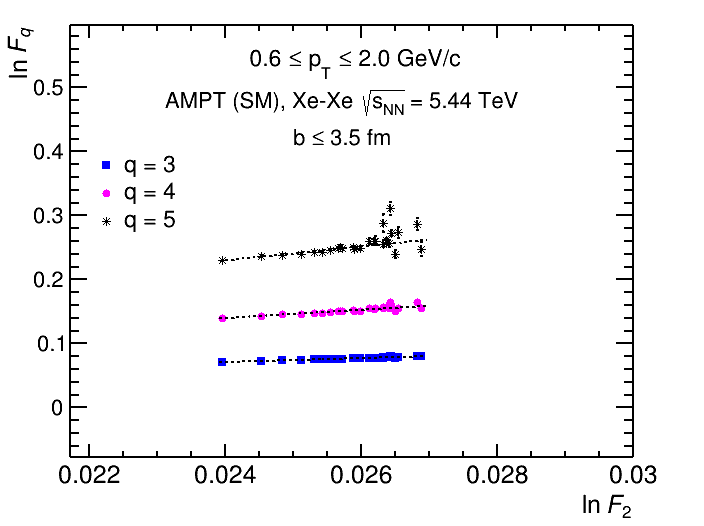}
 	\caption{$ln\, \fq$ for $q$ = 3, 4 and 5 as a function of  $ln\,\fqtwo$ for the charged particles generated in the $\pt$ interval $0.6 \leq p_{\rm{T}} \leq 2.0$~GeV/\textit{c} shows a weak linear behaviour. Lines connecting the data points are the line fits to obtain the slope $\beta_{q}$.}
 	\label{fig:figure7}
    \end{figure}
 \begin{figure}[h]
  	\centering
  	\includegraphics[width=8.cm, height=6.2cm]{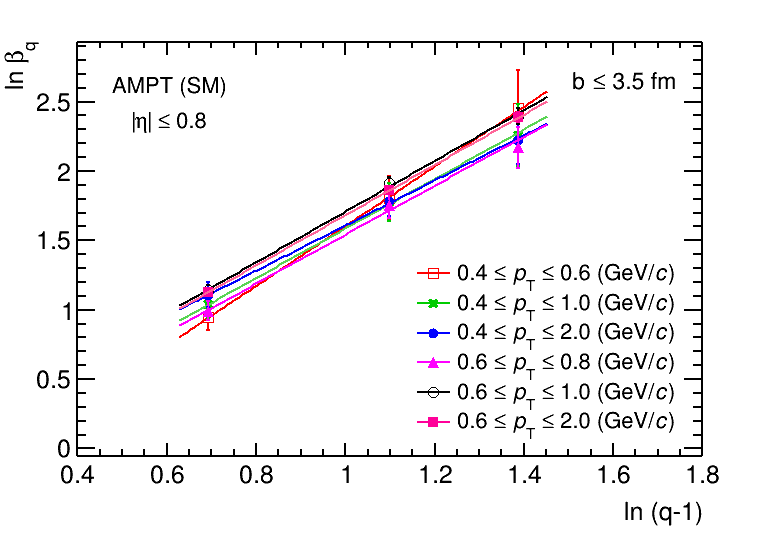}
  	\caption{$ln\beta_q$  vs  $ln(q-1)$ for the different $p_{\rm{T}}$ intervals. Lines connecting the data points are the line fits performed to obtain the scaling exponent $\nu$.}
  	\label{fig:figure9}
  \end{figure}
\begin{figure}[t]
\centering
\includegraphics[width=8.cm,height=6.2cm]{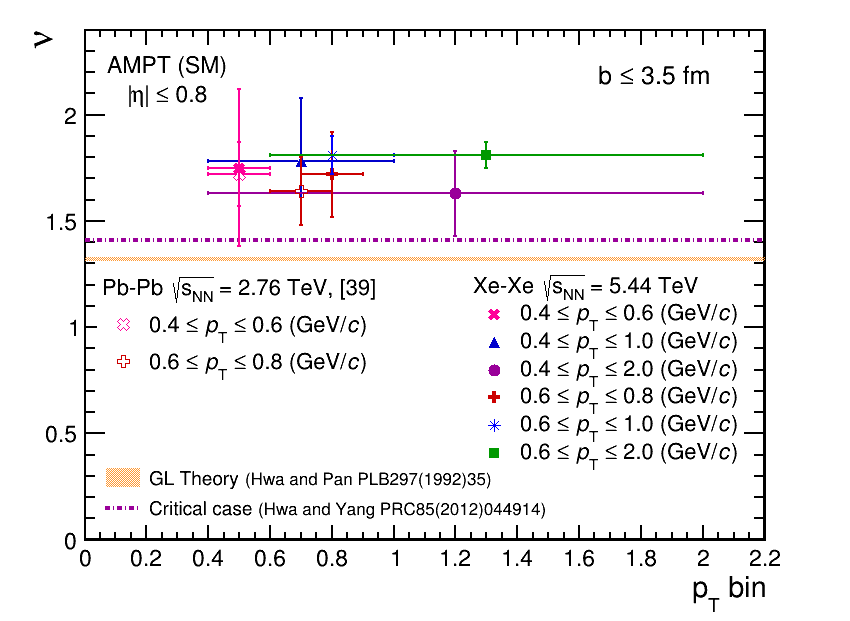}
\caption{The scaling exponent $\nu$ as a function of $p_{\rm{T}}$ for Xe--Xe collision at $\sqrtSnn$ and its comparison with the Pb-Pb system at 2.76 TeV~\cite{Sharma:2018vtf}. Horizontal lines on the markers show the width of the $p_{\rm{T}}$ bin and the vertical lines are the fitting errors.}
   \label{fig:figure11}
  \end{figure}

To quantify fluctuations present in the generated particles, order scaling that is $F_{\mathrm{q}}(M)$'s dependence on $F_{\mathrm{2}}(M)$ is investigated (Eq.\eqref{eqn:fscaling}). Fig.~\ref{fig:figure7} shows the $\ln F_{\mathrm{q}}$ for $q = 3, 4, 5$ as a function of $\ln F_{\mathrm{2}}$ in case for charged particles generated in $p_{\mathrm{T}}$ bin $0.6 \leq p_{\mathrm{T}} \leq 2.0$ GeV/\textit{c}. A weak linear dependence of $\ln F_{\mathrm{q}}$ on $\ln F_{\mathrm{2}}$ is observed. A similar weak F-scaling behaviour is observed for the other $p_{\mathrm{T}}$ bins as well. A line fit performed on these plots gives the slope $\beta_{\mathrm{q}}$. Fig.~\ref{fig:figure9} shows the $\ln \beta_{\mathrm{q}}$ versus $\ln (q - 1)$ graphs, the line fits performed on which give the scaling exponent ($\nu$). A summary of $\nu$ values from all the transverse momentum bins studied here is shown in Fig.~\ref{fig:figure11}. For all $p_{\mathrm{T}}$ bins $\nu$ is found to have a value $\simeq 1.78\pm 0.04$ which is different from the value predicted by the model~\cite{Yang:2017aaz} with critical fluctuations in the spatial configurations of the system  or a second order phase transition~\cite{Hwa:1992ii,Ochs:1990wa,Hwa:1992cn,Cao:1996tg}. The $\nu$ values obtained here quantify the  baseline fluctuations in the system formed in the collisions of deformed nuclei in the SM mode of AMPT model.
\section{Summary}
To summarize, the intermittency analysis is  performed for the charged particles generated in Xe--Xe collisions at $\sqrtSnn$ using the string melting mode of the AMPT model. The charged particles  generated in the central events with impact parameter b $\leq$ 3.5$~$fm in different $\pt$ intervals with $|\eta| \leq 0.8$ in full azimuth and $p_{\rm{T}}\leq 2.0$ GeV/c are investigated. Normalized factorial moments are calculated for the moment order $q$ = 2, 3, 4 and 5 which do not show significant power-law behaviour with number of bins (M) in the phase space but a weak dependence of higher order NFM on the second order NFM (F-scaling) is observed. The average value of the scaling exponent $\nu$ is found to be significantly greater than the value anticipated by the Ginzburg-Landau theory for the second-order phase transition that is 1.304 and that predicted by models with critical fluctuations. The dynamics of critical fluctuations is not incorporated in the AMPT and the observations on $\nu$ made here confirm this and hence supporting that the methodology is suitable to search for the critical point of nuclear matter phase diagram.  Further, the study of fractal parameters of the system generated in these collisions with  the SM AMPT, $D_\text{q}$ as a function of the order of the moments $q$, shows a monofractal nature in particle generation and hence the absence of multifractal complex particle generation. No minimum in the value of  $\lambda_{\rm{q}}$ is observed for $q$ upto 5.  The absence of nontrivial $D_\text{q}$  dependence  on $q$ and the deviation of scaling exponents from critical expectations indicate that fluctuations in the SM AMPT remain governed by smooth transport dynamics rather than critical behaviour.  Similar trends have earlier been reported for symmetric heavy-ion systems such as Pb–Pb and Au–Au using the AMPT string-melting mode. 
The present analysis suggests scaling behaviours of multiplicity fluctuations have a weak dependence on system size or nuclear deformation. Thus within the AMPT model framework, Xe–Xe collisions, despite constituting an intermediate-sized and geometrically deformed nuclear system do not exhibit signatures of intermittency, multifractality, or phase-transition–like scaling in multiplicity fluctuations. 

\section*{Acknowledgments}
  R. Gupta is thankful to  RUSA 2.0 grant to the University of Jammu by the Ministry of Education, Govt. of India, which is utilized for partial support for computing resources. Z. Banoo is thankful to the UGC for granting the JRF/SRF. The authors would like to acknowledge the services provided by the DAE-DST funded WLCG-GRID computing facility at VECC-Kolkata, India, to facilitate the performance of a part of the computation used in this work.

\end{document}